\documentstyle[prabib,aps]{revtex}
\begin{document}
\twocolumn

\wideabs{
\title{\bf Some general bounds for 1-D scattering.} 
\author{Matt Visser$^{\dagger}$}
\address{Physics Department, Washington University, 
Saint Louis MO 63130-4899, USA}
\date{1 October 1998; \LaTeX-ed \today}

%------------------
\maketitle
%-------------------

%------------------------------------------------------------------------------
%Abstract
%------------------------------------------------------------------------------

{\small One-dimensional scattering problems are of wide physical
interest and are encountered in many diverse applications. In this
article I establish some very general bounds for reflection and
transmission coefficients for one-dimensional potential
scattering. Equivalently, these results may be phrased as general
bounds on the Bogolubov coefficients, or statements about the transfer
matrix. A similar analysis can be provided for the parametric change
of frequency of a harmonic oscillator. A number of specific examples
are discussed---in particular I provide a general proof that sharp
step function potentials always scatter more effectively than the
corresponding smoothed potentials. The analysis also serves to collect
together and unify what would otherwise appear to be quite unrelated
results.\\
Physical Review {\bf A59} (1999) 427--438.}

\pacs{} 

}

%---------------------------------------------------------------------
\section{Introduction}
%---------------------------------------------------------------------
\def\Im{\hbox{Im}}
\def\Re{\hbox{Re}}
\def\sech{\hbox{sech}}
\def\max{\hbox{max}}
\def\min{\hbox{min}}
\def\half{{1\over2}}
\def\define{\equiv}
%---------------------------------------------------------------------

One-dimensional scattering problems occur in a wide variety of
physical contexts. In acoustics one might be interested in the
propagation of sound waves down a long pipe, while in electromagnetism
one might be interested in the physics of wave-guides. In quantum
physics the canonical examples are barrier penetration and reflection,
while in classical physics an equivalent problem is the analysis of
parametric resonances. All of these physical problems can be analyzed
in the same mathematical framework, though for definiteness I shall
present the discussion in terms of the Schrodinger equation,
commenting on alternative formulations as appropriate.

For one-dimensional scattering problems there is a large catalog of
specific potentials for which exact analytic results are known. There
are also well-developed numerical techniques for estimating the
scattering properties. In this paper I wish to take a different tack:
I shall develop a number of very general and rather simple {\em
bounds} on the reflection and transmission probabilities
(equivalently, these bounds can be presented in terms of the Bogolubov
coefficients, or in terms of statements about the transfer
matrix). These bounds, because they are so general, are powerful aids
in the {\em qualitative} understanding of one-dimensional
scattering. Furthermore, this analysis provides a unifying theme that
serves to connect together seemingly quite disparate results obtained
in individual special cases.

%---------------------------------------------------------------------
\section{General Analysis}
%---------------------------------------------------------------------
\subsection{Shabat--Zakharov systems}
%---------------------------------------------------------------------

Consider the one-dimensional time-independent Schrodinger equation
[1--15]
\begin{equation}
\label{E:SDE}
-{\hbar^2\over2m} {d^2\over dx^2} \psi(x) + V(x) \; \psi(x) = E \; \psi(x).
\end{equation}
If the potential asymptotes to a constant,
\begin{equation}
V(x\to\pm\infty) \to
V_{\pm\infty},
\end{equation}
then in each of the two asymptotic regions there are two
independent solutions to the Schrodinger equation
\begin{equation}
\psi(\pm i;\pm\infty;x) \approx 
{\exp(\pm i k_{\pm\infty} x) \over \sqrt{k_{\pm\infty}}}.
\end{equation}
Here the $\pm i$ distinguishes right-moving modes $e^{+ikx}$ from
left-moving modes $e^{-ikx}$, while the $\pm \infty$ specifies which
of the asymptotic regions we are in. Furthermore
\begin{equation}
k_{\pm\infty} = {\sqrt{2m(E-V_{\pm\infty})}\over\hbar}.
\end{equation}
To even begin to set up a scattering problem the minimum requirements
are that the potential asymptote to some constant, and this assumption
will be made henceforth.

The so-called Jost solutions~\cite{Chadan-Sabatier} are exact
solutions $J_\pm(x)$ of the Schrodinger equation that satisfy
\begin{equation}
J_+(x\to +\infty) \to {\exp( +i k_{+\infty} x) \over \sqrt{k_{+\infty}}},
\end{equation}
\begin{equation}
J_-(x\to -\infty) \to {\exp( -i k_{-\infty} x) \over \sqrt{k_{-\infty}}},
\end{equation}
and
\begin{equation}
J_+(x\to -\infty) \to 
\alpha {\exp( +i k_{-\infty} x) \over \sqrt{k_{-\infty}}} +
\beta  {\exp( -i k_{-\infty} x) \over \sqrt{k_{-\infty}}},
\end{equation}
\begin{equation}
J_-(x\to +\infty) \to 
\alpha^* {\exp( -i k_{+\infty} x) \over \sqrt{k_{+\infty}}} +
\beta^*  {\exp( +i k_{+\infty} x) \over \sqrt{k_{+\infty}}}.
\end{equation}
Here $\alpha$ and $\beta$ are the (right-moving) Bogolubov
coefficients, which are related to the (right-moving)
reflection and transmission amplitudes by
\begin{equation}
r = {\beta\over\alpha}; \qquad t = {1\over\alpha}.
\end{equation}
These conventions correspond to an incoming flux of right-moving
particles (incident from the left) being partially transmitted and
partially scattered.  The left-moving Bogolubov coefficients are just
the complex conjugates of the right-moving coefficients, however it
should be borne in mind that the phases of $\beta$ and $\beta^*$ are
physically meaningless in that they can be arbitrarily changed simply
by moving the origin of coordinates. The phases of $\alpha$ and
$\alpha^*$ on the other hand do contain real physical information.

In this article I will derive some very general bounds on $|\alpha|$
and $|\beta|$, which also lead to general bounds on the reflection and
transmission probabilities
\begin{equation}
R = |r|^2$; \qquad  $T=|t|^2.
\end{equation}
The key idea is to re-write the second-order Schrodinger equation as a
particular type of Shabat-Zakharov~\cite{Eckhaus-van-Harten} system: a
particular set of two coupled first-order differential equations for
which bounds can be easily established. A similar representation of
the Schrodinger equation is briefly discussed by Peirls~\cite{Peirls}
and related representations are well-known, often being used used
without giving an explicit reference (see {\em e.g.} \cite{Lindig}).
However an exhaustive search has not uncovered prior use of the
particular representation of this article, nor the idea of using the
representation to place bounds on one-dimensional scattering.

I start by introducing an arbitrary auxiliary function $\varphi(x)$
which may be either real or complex, though I do demand that
$\varphi'(x)\neq 0$, and then defining
\begin{equation}
\label{E:representation}
\psi(x) = 
a(x) {\exp(+i \varphi)\over\sqrt{\varphi'}} + 
b(x) {\exp(-i \varphi)\over\sqrt{\varphi'}}.
\end{equation}
This representation effectively seeks to use quantities resembling the
``phase integral'' wavefunctions as a basis for the true
wavefunction~\cite{Froman-Froman}. This representation is of course
highly redundant, since one complex number $\psi(x)$ has been traded
for two complex numbers $a(x)$ and $b(x)$ plus an essentially
arbitrary auxiliary function $\varphi(x)$. In order for this
representation to be most useful it is best to arrange things so that
$a(x)$ and $b(x)$ asymptote to constants at spatial infinity, which we
shall soon see implies that we should pick the auxiliary function to
satisfy 
\begin{equation}
\varphi'(x) \to k_{\pm\infty} \qquad \hbox{as} \qquad x\to\pm\infty. 
\end{equation}
To trim down the number of degrees of freedom it is useful to impose a
``gauge condition''
\begin{equation}
\label{E:gauge}
{d\over dx}\left({a\over\sqrt{\varphi'}}\right) e^{+i \varphi} + 
{d\over dx}\left({b\over\sqrt{\varphi'}}\right) e^{-i \varphi} = 0.
\end{equation}
Subject to this gauge condition,
\begin{equation}
\label{E:gradient}
{d\psi\over dx} = i \sqrt{\varphi'} 
\left\{ a(x) \exp(+i \varphi) - b(x) \exp(-i \varphi ) \right\}.
\end{equation}
I now re-write the Schrodinger equation in terms of two coupled
first-order differential equations for these position-dependent
Bogolubov coefficients. To do this note that
\begin{eqnarray}
\label{E:double-gradient}
{d^2\psi\over dx^2} 
&=& 
{d\over dx} 
\left( i{\varphi'\over \sqrt{\varphi'}} 
\left\{ a e^{+i \varphi} - b e^{-i \varphi} \right\} 
\right)
\\
&=& {(i\varphi')^2\over\sqrt{\varphi'}} 
\left\{ a e^{+i \varphi} + b e^{-i \varphi} \right\}
\nonumber\\
&+&  i \varphi'
\left\{ 
{d\over dx}\left({a\over\sqrt{\varphi'}}\right) e^{+i \varphi} - 
{d\over dx}\left({b\over\sqrt{\varphi'}}\right) e^{-i \varphi} 
\right\}
\nonumber\\
&+&  i{\varphi''\over\sqrt{\varphi'}} 
\left\{ a e^{+i \varphi} - b e^{-i \varphi} \right\}
\\
&=& -{\varphi'^2\over\sqrt{\varphi'}}  
\left\{ a e^{+i \varphi} + b e^{-i \varphi} \right\}
\nonumber\\
&+&  {2 i\varphi'\over\sqrt{\varphi'}}  {da\over dx} e^{+i \varphi} -
i{\varphi''\over\sqrt{\varphi'}} b e^{-i \varphi} 
\\
&=& -{\varphi'^2\over\sqrt{\varphi'}}
  \left\{ a e^{+i \varphi} + b e^{-i \varphi} \right\}
\nonumber\\
&-&  {2 i\varphi'\over\sqrt{\varphi'}}  {db\over dx} e^{-i \varphi} + 
i{\varphi''\over\sqrt{\varphi'}} a e^{+i \varphi}.
\end{eqnarray}
(The last two relations use the ``gauge condition''.) Now insert
these formulae into the Schrodinger equation written in the form
\begin{equation}
{d^2\psi\over dx^2} = - k(x)^2 \; \psi \equiv
- {2m(E-V(x))\over\hbar^2} \; \psi,
\end{equation}
to deduce
\begin{eqnarray}
\label{E:system-a}
{da\over dx} &=& +
{1\over 2\varphi'} 
\Bigg\{ 
\varphi'' \; b \; \exp(-2i\varphi) 
\nonumber\\
&&\qquad
+ 
i\left[k^2(x)-(\varphi')^2\right] \left( a + b \exp(-2i\varphi) \right)
\Bigg\},
\\
\label{E:system-b}
{db\over dx} &=& +
{1\over 2\varphi'} 
\Bigg\{ 
\varphi'' \; a \; \exp(+2i\varphi) 
\nonumber\\
&&\qquad
- i\left[k^2(x)-(\varphi')^2\right] \left( b + a \exp(+2i\varphi) \right)
\Bigg\}.
\end{eqnarray}
It is easy to verify that this first-order system is compatible with
the ``gauge condition'' (\ref{E:gauge}), and that by iterating the
system twice (subject to this gauge condition) one recovers exactly
the original Schrodinger equation. These equations hold for arbitrary
$\varphi$, real or complex, and when written in matrix form, exhibit a
deep connection with the transfer matrix
formalism~\cite{Transfer-matrix}.

%------------------------------------------------------------------------------
\subsection{Bounds} 
%-----------------------------------------------------------------------------

To obtain our bounds on the Bogolubov coefficients we start by
restricting attention to the case that $\varphi(x)$ is a {\em real}
function of $x$.  (Since $\varphi$ is an essentially arbitrary
auxiliary function this is not a particularly restrictive
condition). Under this assumption the probability current is
\begin{equation}
{\cal J} 
= \Im\left\{ \psi^* {d\psi \over dx} \right\} 
= \left\{ |a|^2 - |b|^2 \right\}.
\end{equation}
Now at $x\sim+\infty$ the wavefunction is purely right-moving and
normalized to 1, because we are considering one-dimensional Jost
solutions~\cite{Chadan-Sabatier}. Then for all $x$ we have a conserved
quantity
\begin{equation}
\label{E:conservation}
|a|^2 - |b|^2  = 1.
\end{equation}
{\em It is this result that makes it useful to interpret $a(x)$ and
$b(x)$ as position-dependent Bogolubov coefficients relative to the
auxiliary function $\varphi(x)$}.  Now use the fact that
\begin{equation}
{d|a|\over dx} = 
{1\over2|a|} \left( a^* {da\over dx} + a {da^* \over dx} \right),
\end{equation}
and use equation (\ref{E:system-a}) to obtain
\begin{eqnarray}
{d|a|\over dx} &=& 
{1\over2|a|} {1\over2\varphi'}
\Big( 
\varphi'' 
\left[a^* b \exp(-2i\varphi) + a b^* \exp(+2i\varphi) \right]
\nonumber\\
&&
+ i[k^2 -(\varphi')^2] 
\left[a^* b \exp(-2i\varphi) - a b^* \exp(+2i\varphi) \right]
\Big).
\nonumber\\
&&
\end{eqnarray}
That is
\begin{eqnarray}
{d|a|\over dx} &=& 
{1\over2|a|} {1\over2\varphi'}
\Re\Bigg( 
\left[\varphi'' + i[k^2 -(\varphi')^2] \right] \;
\nonumber\\
&&
\qquad
\left[a^* b \exp(-2i\varphi) \right]
\Bigg).
\nonumber\\
&&
\end{eqnarray}
The right hand side can now be bounded from above, by systematically
using $\Re(A\;B) \leq |A| \; |B|$. This leads to
\begin{equation}
{d |a|\over dx} \leq  
{\sqrt{(\varphi'')^2+ \left[k^2-(\varphi')^2\right]^2} \over 2 |\varphi'| } 
\; |b|.
\end{equation}
It is essential that $\varphi$ be real to have $|\exp(-2i\varphi)|=1$
which is the other key ingredient above.  Now define the non-negative
quantity
\begin{equation}
\label{E:vartheta}
\vartheta[\varphi(x),k(x)] \define 
{\sqrt{(\varphi'')^2+ \left[k^2(x)-(\varphi')^2\right]^2} 
\over 2 |\varphi'| }, 
\end{equation}
and use the conservation law (\ref{E:conservation}) to write
\begin{equation}
{d |a|\over dx} \leq  \vartheta \sqrt{|a|^2 -1}.
\end{equation}
Integrate this inequality
\begin{equation}
\left. \left\{\cosh^{-1} |a| \right\} \right|_{x_i}^{x_f} \leq  
\int_{x_i}^{x_f}  \vartheta \; dx.
\end{equation}
Taking limits as $x_i\to-\infty$ and $x_f\to+\infty$
\begin{equation}
\cosh^{-1} |\alpha| \leq  
\int_{-\infty}^{+\infty}  \vartheta \; dx.
\end{equation}
That is
\begin{equation}
\label{B:alpha0}
|\alpha| \leq  
\cosh\left(  \int_{-\infty}^{+\infty}  \vartheta \; dx \right).
\end{equation}
Which automatically implies
\begin{equation}
\label{B:beta0}
|\beta| \leq  
\sinh\left(   \int_{-\infty}^{+\infty}  \vartheta \; dx \right).
\end{equation}
Since this result holds for {\em all real} choices of the auxiliary
function $\varphi(x)$, (subject only to $\varphi' \neq 0$ and
$\varphi' \to k_{\pm\infty}$ as $x \to \pm\infty$), it encodes an
enormously wide class of bounds on the Bogolubov coefficients. When
translated to reflection and transmission coefficients the equivalent
statements are
\begin{equation}
\label{B:T0}
T \geq 
\sech^2
\left(
\int_{-\infty}^{+\infty} \vartheta \; dx
\right),
\end{equation}
and
\begin{equation}
\label{B:R0}
R \leq 
\tanh^2
\left(
\int_{-\infty}^{+\infty} \vartheta \; dx
\right).
\end{equation}
I shall soon turn this general result into more specific theorems.

%------------------------------------------------------------------------------
\subsection{Transfer matrix representation} 
%-----------------------------------------------------------------------------

The system of equations (\ref{E:system-a})--(\ref{E:system-b}) can
also be written in matrix form. It is convenient to define
\begin{equation}
\rho \define \varphi'' +i[k^2(x)-(\varphi')^2].
\end{equation}
Then
\begin{equation}
{d\over dx} \left[\matrix{ a \cr b}\right] = 
{1\over2\varphi'} 
\left[
\matrix{ i \Im[\rho] & 
\rho\exp(-2i\varphi) \cr 
\rho^*\exp(+2i\varphi)  & 
-i \Im[\rho]}
\right]
\left[ \matrix{ a \cr b}\right].
\end{equation}
This has the formal solution
\begin{equation}
\left[\matrix{ a(x_f) \cr b(x_f)}\right] = 
E(x_f,x_i)  \left[\matrix{ a(x_i) \cr b(x_i)}\right],
\end{equation}
in terms of a generalized position-dependent ``transfer
matrix''~\cite{Transfer-matrix}
\begin{eqnarray}
&&
E(x_f,x_i) = 
\nonumber\\
&&
{\cal P} \exp\left( 
\int_{x_i}^{x_f}
{1\over2\varphi'} 
\left[
\matrix{ i \Im[\rho] & 
\rho\exp(-2i\varphi) \cr 
\rho^*\exp(+2i\varphi)  & 
-i \Im[\rho]}
\right] 
 dx \right),
\nonumber\\
&&
\end{eqnarray}
where the symbol ${\cal P}$ denotes ``path ordering''. In particular,
if we take $x_i\to-\infty$ and $x_f\to+\infty$ we obtain a formal but
exact expression for the Bogolubov coefficients
\begin{eqnarray}
&&
\left[ \matrix{ \alpha & \beta^* \cr \beta & \alpha^*} \right] =
E(\infty,-\infty) = 
\nonumber\\
&&
{\cal P} \exp\left( 
\int_{-\infty}^\infty
{1\over2\varphi'} 
\left[
\matrix{ i \Im[\rho] & 
\rho\exp(-2i\varphi) \cr 
\rho^*\exp(+2i\varphi)  & 
-i \Im[\rho]}
\right] 
 dx \right).
\nonumber\\
&& 
\end{eqnarray}
The matrix $E$ is {\em not} unitary, though it does have determinant
1. It is in fact an element of the group $SU(1,1)$.  Taking
\begin{equation}
\sigma_z =  \left[ \matrix{ +1 & 0 \cr 0 & -1} \right],
\end{equation}
then $(\sigma_z)^2 = +I$, and defining $E^\dagger = (E^*)^T$, it is
easy to see
\begin{equation}
E^\dagger \sigma_z E = \sigma_z.
\end{equation}
This is the analog of the invariance of the Minkowski metric for
Lorentz transformations in $SO(3,1)$.  Similarly, if we define the
``complex structure'' $J$ by
\begin{equation}
J =  \left[ \matrix{ 0 & 1 \cr -1 & 0} \right],
\end{equation}
then $J^2 = -I$ and
\begin{equation}
E^\dagger =  J E J.
\end{equation}
%

%------------------------------------------------------------------------------
\section{Special Case 1} 
%-----------------------------------------------------------------------------

Suppose now that the potential satisfies $V_{+\infty} =
V_{-\infty}$. Also, choose the phase function $\varphi(x)$ to be
$\varphi= k_\infty \, x$. We also require $k_\infty \neq 0$, that is
$E > V_{\pm \infty}$.  This is the special case discussed in a
different context by Peirls~\cite{Peirls}.  Then the evolution
equations simplify tremendously, and
\begin{equation}
\vartheta 
\to {|k^2 - k_\infty^2| \over 2 k_\infty} 
= {m |V(x)-V_\infty|\over \hbar^2 k_\infty}.
\end{equation}
Using $(\hbar k_\infty)^2=2m(E-V_\infty)$, the bounds become
\begin{equation}
\label{B:T1}
T \geq 
\sech^2
\left(
{1\over\hbar}\sqrt{m\over2(E-V_\infty)}
\int_{-\infty}^{+\infty} |V-V_\infty| \; dx
\right),
\end{equation}
and
\begin{equation}
\label{B:R1}
R \leq 
\tanh^2
\left(
{1\over\hbar}\sqrt{m\over2(E-V_\infty)}
\int_{-\infty}^{+\infty} |V-V_\infty| \; dx
\right).
\end{equation}
These bounds are exact non-perturbative results, however for high
energies it may be convenient to use the slightly less restrictive
(but analytically much more tractable) bounds
\begin{equation}
\label{B:T1-weak}
T \geq 
1 - {m 
\left(\int_{-\infty}^{+\infty} |V-V_\infty| dx\right)^2 
\over 2E\hbar^2},
\end{equation}
and
\begin{equation}
\label{B:R1-weak}
R \leq 
{m 
\left(\int_{-\infty}^{+\infty} |V-V_\infty| dx\right)^2 
\over 2E\hbar^2}.
\end{equation}
This version of the bounds also holds for all energies, but is not
very restrictive for low energy.

The transfer matrices can be analyzed by checking that the evolution
equations simplify to
\begin{eqnarray}
\label{E:system1a}
{da\over dx} &=&
{-im(V-V_\infty)\over\hbar^2 k_\infty}
\left\{ a + b \exp(-2ik_\infty x) \right\},
\\
\label{E:system1b}
{db\over dx} &=&
{+im(V-V_\infty)\over\hbar^2 k_\infty}
\left\{ a \exp(+2ik_\infty x) + b \right\}.
\end{eqnarray}
This can be written in matrix form as
\begin{eqnarray}
{d\over dx} \left[\matrix{ a \cr b}\right] &=& 
{-im(V-V_\infty)\over\hbar^2 k_\infty} 
\nonumber\\
&&
%\qquad
\left[\matrix{ 1 & \exp(-2ik_\infty x) \cr 
     -\exp(+2ik_\infty x) & -1}\right]
\left[ \matrix{ a \cr b}\right].
\nonumber\\
&&
\end{eqnarray}
This version of the Shabat--Zakharov system~\cite{Eckhaus-van-Harten}
has a formal solution in terms of the transfer matrix
\begin{eqnarray}
&&
E(x_f,x_i) = 
\nonumber\\
&&
{\cal P} \exp\Bigg( {-im\over\hbar^2 k_\infty} 
\int_{x_f}^{x_i} (V(x)-V_\infty) 
\nonumber\\
&&
\qquad\qquad
\left[\matrix{ 1 & e^{-2ik_\infty x} \cr 
     -e^{+2ik_\infty x} & -1} \right] dx \Bigg),
\end{eqnarray}
The formal but exact expression for the Bogolubov coefficients is now
\begin{eqnarray}
&&
\left[ \matrix{ \alpha & \beta^* \cr \beta & \alpha^*} \right] =
E(\infty,-\infty) = 
\nonumber\\
&&
{\cal P} \exp\Bigg( {-im\over\hbar^2 k_\infty} 
\int_{-\infty}^\infty (V(x)-V_\infty) 
\nonumber\\
&&
\qquad\qquad
\left[ \matrix{ 1 & e^{-2ik_\infty x} \cr 
     -e^{+2ik_\infty x} & -1} \right] dx \Bigg).
\end{eqnarray}
Furthermore, the form of the system
(\ref{E:system1a})--(\ref{E:system1b}) suggests that it might be
useful to define
\begin{eqnarray}
a &=& 
\tilde a \; \exp\left[+{im\over\hbar^2 k_\infty} 
\int_{-\infty}^x (V(y)-V_\infty) dy\right],
\\
b &=& 
\tilde b \; \exp\left[-{im\over\hbar^2 k_\infty} 
\int_{-\infty}^x (V(y)-V_\infty) dy\right].
\end{eqnarray}
Then
\begin{eqnarray}
\label{E:system1ba}
{d\tilde a\over dx} &=&
{-im(V(x)-V_\infty)\over\hbar^2 k_\infty}
\; \tilde b \; \exp(-2ik_\infty x),
\\
\label{E:system1bb}
{d\tilde b\over dx} &=&
{+im (V(x)-V_\infty)\over\hbar^2 k_\infty}
\; \tilde a \; \exp(+2ik_\infty x).
\end{eqnarray}
This representation simplifies some of the results, for instance
\begin{eqnarray}
&&
\left[ 
\matrix{ \tilde\alpha &  \tilde\beta^* \cr 
         \tilde\beta & \tilde\alpha^*} \right] =
\tilde E(\infty,-\infty) = 
\nonumber\\
&&
{\cal P} \exp\Bigg( {-im\over\hbar^2 k_\infty} 
\int_{-\infty}^\infty (V(x)-V_\infty) 
\nonumber\\
&&
\qquad\qquad
\left[ \matrix{ 0 & e^{-2ik_\infty x} \cr 
     -e^{+2ik_\infty x} & 0} \right] dx \Bigg).
\end{eqnarray}
This can be used as the basis of an approximation scheme for
$\tilde\beta$. Suppose that for all $x$ we have $|\tilde b(x)| \ll 1$,
so that $\tilde a(x)| \approx 1$. Then
\begin{equation}
{d\tilde b\over dx} \approx
{+im(V(x)-V_\infty)\over\hbar^2 k_\infty}
\; \exp(+2ik_\infty x).
\end{equation}
This may be immediately integrated to yield
\begin{equation}
\tilde\beta \approx 
{+im\over\hbar^2 k_\infty} \int_{-\infty}^{+\infty} (V(x)-V_\infty) 
\; \exp(+2ik_\infty x) \; dx.
\end{equation}
This is immediately recognizable as the (first) Born approximation. If
we instead work in terms of the original definition $\beta$
\begin{eqnarray}
\beta 
&\approx& 
{+im\over\hbar^2 k_\infty} 
\exp\left[+{im\over\hbar^2 k_\infty} 
\int_{-\infty}^{+\infty} (V(x)-V_\infty) \; dx \right]
\nonumber\\
&&
\times
\int_{-\infty}^{+\infty}  (V(x)-V_\infty) 
\; \exp(+2ik_\infty x)
\nonumber\\
&&
\qquad\qquad  
\exp\left[-{im\over\hbar^2 k_\infty} 
\int_{-\infty}^x (V(y)-V_\infty) \; dy\right] dx.
\nonumber\\
\end{eqnarray}
This is one form of the distorted Born wave approximation.

In short, this type of analysis collects together a large number of
results that otherwise appear quite unrelated. By taking further
specific cases of these bounds and related results it is possible to
reproduce many analytically known results, such as for delta function
potentials, double delta function potentials, square wells, and
$\sech^2$ potentials, as discussed later in this article. (See Section
VI.)

%------------------------------------------------------------------------------
\section{Special Case 2} 
%-----------------------------------------------------------------------------

Suppose now we take $k(x) = \varphi'(x)$. This means that we are
choosing our auxiliary function so that we use the WKB approximation
for the true wavefunction as a ``basis'' for calculating the Bogolubov
coefficients. This choice is perfectly capable of handling the case
$V_{+\infty}\neq V_{-\infty}$ but because of the assumed reality of
$\varphi$ is limited to considering scattering {\em over} the
potential barrier.  (This is the special case implicit in a different
context in~\cite{Lindig}). The evolution equations again simplify
tremendously, to yield
\begin{eqnarray}
\label{E:system-2a}
{da\over dx} &=& +
{1\over 2\varphi'} 
\Bigg\{ 
\varphi'' \; b \; \exp(-2i\varphi)
\Bigg\}, 
\\
\label{E:system-2b}
{db\over dx} &=& +
{1\over 2\varphi'} 
\Bigg\{ 
\varphi'' \; a \; \exp(+2i\varphi) 
\Bigg\}.
\end{eqnarray}
This form of the evolution equations can be related to the qualitative
discussion of scattering over a potential barrier presented by
Migdal~\cite{Migdal1,Migdal2}.  For this choice of auxiliary function
\begin{equation}
\vartheta \to {|\varphi''|\over2|\varphi'|} = {|k'|\over2|k|},
\end{equation}
and the bounds become
\begin{equation}
\label{B:T2}
T \geq  
\sech^2\left( {1\over2} \int_{-\infty}^{+\infty}  {|k'|\over |k|} dx \right),
\end{equation}
and
\begin{equation}
\label{B:R2}
R \leq  
\tanh^2\left(  {1\over2} \int_{-\infty}^{+\infty}  {|k'|\over |k|} dx \right).
\end{equation}
The relevant transfer matrix is now
\begin{eqnarray}
&&
E(x_f,x_i) = 
\nonumber\\
&&
{\cal P} \exp\left( \half
\int_{x_f}^{x_i} {\varphi''\over\varphi'}
\left[\matrix{0 & e^{-2i\varphi} \cr e^{+2i\varphi} & 0} \right] dx \right),
\end{eqnarray}
The Bogolubov coefficients are now
\begin{eqnarray}
&&
\left[ \matrix{ \alpha & \beta^* \cr \beta & \alpha^*} \right] =
E(\infty,-\infty) = 
\nonumber\\
&&
{\cal P} \exp\left( 
\int_{-\infty}^\infty {\varphi''\over\varphi'}
\left[ \matrix{ 0 & e^{-2i\varphi} \cr e^{+2i\varphi} & 0} \right] dx \right).
\end{eqnarray}
This type of analysis collects together and unifies several
analytically known results for scattering over the barrier, such as
for asymmetric square wells and Poschl-Teller potentials. (See Section
VI.) After a few general comments, I shall turn to specializing this
still rather general result to more specific cases.

%------------------------------------------------------------------------------
\subsection{Reflection above the barrier}
%------------------------------------------------------------------------------

The system (\ref{E:system-2a})--(\ref{E:system-2b}) can also be used
as the basis of an approximation scheme for $\beta$. Suppose that for
all $x$ we have $|b(x)| \ll 1$, so that $|a(x)| \approx 1$. Then
\begin{equation}
{d b\over dx} \approx
{\varphi''\over2 \varphi'} \;
\exp(+2i\varphi).
\end{equation}
This may be immediately integrated to yield
\begin{equation}
\beta \approx 
{1\over2} \int_{-\infty}^{+\infty} 
{\varphi''(x)\over \varphi'(x)} \;
\exp\left(+2i\varphi\right) \; dx.
\end{equation}
Or the equivalent
\begin{equation}
\beta \approx 
{1\over2} \int_{-\infty}^{+\infty} 
{k'(x)\over k(x)} \;
\exp\left(+2i\int_{-\infty}^x k(y) dy\right) \; dx.
\end{equation}
This result serves to clarify the otherwise quite mysterious
discussion of ``reflection above the barrier'' given by
Migdal~\cite{Migdal1,Migdal2}.  Even though the WKB wavefunctions are
buried in the representation of the wavefunction underlying the
analysis leading to this approximation, the validity of this result
for $|\beta|$ does not require validity of the WKB approximation.

If the shifted potential, $V-V_\infty$, is ``small'' we can recover
the Born approximation in the usual manner. In that case $k' \equiv
mV'/(\hbar^2 k) \approx mV'/(\hbar^2 k_\infty)$, while $\exp(2i\int
k)\approx \exp(2ik_\infty x)$.  A single integration by parts then
yields
\begin{equation}
\beta \approx 
-i{m\over\hbar^2 k_\infty} \int_{-\infty}^{+\infty} 
(V(x)-V_\infty) \;
\exp\left(+2ik_\infty x\right) \; dx.
\end{equation}

%------------------------------------------------------------------------------
\subsection{Under the barrier?}
%------------------------------------------------------------------------------

What goes wrong when we try to extend this analysis into the
classically forbidden region?  Analytically continuing the system
(\ref{E:system-2a})--(\ref{E:system-2b}) is trivial, replace
\begin{equation}
\varphi'(x) = k\to i\kappa = i\sqrt{2m(V-E)}/\hbar, 
\end{equation}
and write
\begin{equation}
\varphi(x) = 
\varphi_{\mathrm tp} +i\int_{\mathrm tp}^x \kappa(y) dy,
\end{equation}
to obtain
\begin{eqnarray}
\label{E:system2af}
{da\over dx} &=& +
{\kappa'\over 2\kappa} \; b \; \exp(-2i\varphi_{\mathrm tp}) \;
\exp\left(+2\int \kappa\right),
\\
\label{E:system2bf}
{db\over dx} &=& +
{\kappa'\over 2\kappa} \; a \; \exp(+2i\varphi_{\mathrm tp}) \;
\exp\left(-2\int \kappa\right).
\end{eqnarray}
Thus we are {\em violating} our previous condition that $\varphi$ be
real, though we still require $\varphi'\neq 0$.  This is a perfectly
good Shabat-Zakharov system that works in the forbidden region. But
you cannot now use this to derive bounds on the transmission
coefficient. The fly in the ointment resides in the fact that the
formula for the probability current is modified, and that in the
forbidden region the probability current is
\begin{equation}
{\cal J}  = \Im\left\{ \psi^* {d\psi \over dx} \right\} =  
2\;\Im\left\{ a b^*  \exp(+2i\varphi_{\mathrm tp}) \right\}.
\end{equation}
For a properly normalized flux in the allowed region ($|a|^2-|b|^2=1$),
we have in the forbidden region
\begin{equation}
2\;\Im\left\{ a b^*  \exp(+2i\varphi_{\mathrm tp}) \right\}  = 1.
\end{equation}
While this does imply $2 |a| |b| > 1$, the inequality is unfortunately
in the wrong direction to be useful for placing bounds on the
transmission coefficient.

%------------------------------------------------------------------------------
\subsection{Special Case 2-a} 
%-----------------------------------------------------------------------------

Suppose now that $V(x)$ is continuous and monotonic increasing or
decreasing, varying from $V_{-\infty} = V(-\infty)$ to $V_{+\infty} =
V(+\infty)$.  Suppose $E \geq \max\{V_{-\infty},V_{+\infty}\}$ so
there is no classical turning point. Then
\begin{equation}
\int_{-\infty}^{+\infty}  {|k'|\over|k|} dx = 
\left|\ln\left( {k_{+\infty}\over k_{-\infty}} \right)\right|,
\end{equation}
and the transmission and reflection probabilities satisfy
\begin{equation}
\label{B:T3}
T \geq 
{4 k_{+\infty}k_{-\infty}
\over 
(k_{+\infty}+k_{-\infty})^2
},
\end{equation}
and
\begin{equation}
\label{B:R3}
R \leq 
{(k_{+\infty}-k_{-\infty})^2
\over 
(k_{+\infty}+k_{-\infty})^2}.
\end{equation}
These bounds are immediately recognizable as the exact analytic
results for a step-function
potential~\cite{Landau-Lifshitz,Capri,Stehle}, and the result asserts
that for arbitrary smooth monotonic potentials the step function
provides upper and lower bounds on the exact result. If we are
interested in physical situations such as a time-dependent refractive
index~\cite{Sissa,Yablonovitch}, or particle production due to the
expansion of the universe~\cite{Birrell-Davies}, this technique shows
that {\em sudden} changes in refractive index or size of the universe
provide a strict upper bound on particle production.

%------------------------------------------------------------------------------
\subsection{Special Case 2-b} 
%-----------------------------------------------------------------------------

Suppose now that $V(x)$ has a single unique extremum (either a peak or
a valley), and provided that $E\geq \max\{V_{\infty}, V_{\mathrm
extremum}, V_{+\infty}\}$ so that there is no classical turning point,
then $k(x)$ moves monotonically from $k_{-\infty}$ to $k_{\mathrm
extremum}$ and then back to $k_{+\infty}$. Under these circumstances
\begin{eqnarray}
\int_{-\infty}^{+\infty}  {|k'|\over k} dx 
&=&
\left|\ln\left[{k_{\mathrm extremum}\over k_{-\infty}}\right]\right| +
\left|\ln\left[{k_{\mathrm extremum}\over k_{+\infty}}\right]\right| 
\\
&=&
\left|\ln\left[
{k_{\mathrm extremum}^2\over k_{-\infty} k_{+\infty} }
\right]\right|.
\end{eqnarray}
This implies
\begin{equation}
|\alpha| \leq  
\cosh\left|
\ln\left[{k_{\mathrm extremum}\over \sqrt{k_{-\infty} k_{+\infty}} }\right]
\right|.
\end{equation}
Which yields
\begin{equation}
|\beta| \leq  
\sinh\left|
\ln\left[{k_{\mathrm extremum}\over \sqrt{k_{-\infty} k_{+\infty}} }\right]
\right|.
\end{equation}
To be more specific, if in addition $V(-\infty)=0=V(+\infty)$, so that
$k_{-\infty}=k_{+\infty}$, then we have
\begin{equation}
|\alpha| \leq  
{k_{\mathrm extremum}^2+k_\infty^2\over 
2 k_{\mathrm extremum} k_{\infty} },
\end{equation}
and
\begin{equation}
|\beta| \leq  
{|k_{\mathrm extremum}^2-k_\infty^2|\over 
2 k_{\mathrm extremum} k_{\infty} }.
\end{equation}
Translated into statements about the transmission and reflection
probabilities this becomes
\begin{equation}
\label{B:T2b}
T \geq 
{ (E-V_\infty)(E-V_{\mathrm extremum})
\over 
(E-V_\infty)(E-V_{\mathrm extremum}) + 
{1\over4}(V_{\mathrm extremum}-V_\infty)^2},
\end{equation}
and
\begin{equation}
\label{B:R2b}
R \leq 
{ {1\over4}(V_{\mathrm extremum}-V_\infty)^2
\over 
(E-V_\infty)(E-V_{\mathrm extremum}) + 
{1\over4}(V_{\mathrm extremum}-V_\infty)^2}.
\end{equation}
Equivalently
\begin{equation}
\label{B:T2b*}
T \geq 
 1 - { (V_{\mathrm extremum}-V_\infty)^2
\over 
(2E-V_{\mathrm extremum}-V_\infty)^2},
\end{equation}
and
\begin{equation}
\label{B:R2b*}
R \leq 
{ (V_{\mathrm extremum}-V_\infty)^2
\over 
(2E-V_{\mathrm extremum}-V_\infty)^2}.
\end{equation}
For low energies, these results are weaker than the bounds derived
under Special Case 1, [(\ref{B:T1}), (\ref{B:R1})] and
[(\ref{B:T1-weak}), (\ref{B:R1-weak})], but have the advantage of
requiring more selective information about the potential. For high
energies,
\begin{equation}
E \gg 
{\hbar^2 (V_{\mathrm extremum}-V_\infty)^2 \over 
2 m \left(\int_{-\infty}^{+\infty} |V(x)-V_\infty| dx\right)^2},
\end{equation}
the present result (when it is applicable) leads to tighter bounds on
the transmission and reflection coefficients.

Numerous generalizations of these formulae are possible. For example,
at the cost of a little extra notation, we also already have enough
information to provide a bound on an {\em asymmetric} barrier or {\em
asymmetric} well, as long as it has only a single extremum (maximum or
minimum) we apply the previous equations to derive
\begin{equation}
|\alpha| \leq  
{k_{\mathrm extremum}^2+k_{+\infty}k_{-\infty}
\over
2 k_{\mathrm extremum} \sqrt{k_{+\infty} k_{-\infty}} }.
\end{equation}
and
\begin{equation}
|\beta| \leq  
{|k_{\mathrm extremum}^2-k_{+\infty}k_{-\infty}|
\over
2 k_{\mathrm extremum}  \sqrt{k_{+\infty} k_{-\infty}} }.
\end{equation}
Translated into statements about the transmission
and reflection probabilities this becomes
\begin{equation}
\label{B:T5-1}
T \geq 
{ 
4 k_{+\infty}k_{-\infty} k_{\mathrm extremum}^2
\over 
\left\{  
k_{\mathrm extremum}^2 + k_{+\infty}k_{-\infty}
\right\}^2},
\end{equation}
and
\begin{equation}
\label{B:R5-1}
R \leq 
{
\left\{  
k_{\mathrm extremum}^2 -k_{+\infty}k_{-\infty} 
\right\}^2 
\over 
\left\{  
k_{\mathrm extremum}^2 + k_{+\infty} k_{-\infty}
\right\}^2}.
\end{equation}
Equivalently
\begin{equation}
\label{B:T5}
T \geq 
{ 4(E-V_{\mathrm extremum})\sqrt{(E-V_{+\infty})(E-V_{-\infty})}
\over 
[(E-V_{\mathrm extremum})+\sqrt{(E-V_{+\infty})(E-V_{-\infty})}]^2},
\end{equation}
and
\begin{equation}
\label{B:R5}
R \leq 
{[(E-V_{\mathrm extremum})-\sqrt{(E-V_{+\infty})(E-V_{-\infty})}]^2
\over 
[(E-V_{\mathrm extremum})+\sqrt{(E-V_{+\infty})(E-V_{-\infty})}]^2}.
\end{equation}
This can be compared, for example, with known analytic results for the
asymmetric square well, see equation (\ref{B:asymetric-square-well})
in Section VI.

%------------------------------------------------------------------------------
\subsection{Special Case 2-c} 
%-----------------------------------------------------------------------------

Suppose now that $V(x)$ has a number of extrema, (both peaks and
valleys). I allow $V(+\infty) \neq V(-\infty)$, but demand that for
all extrema $E\geq \max\{V_{-\infty},V_{+\infty},V_{\mathrm
extremum}^i\}$ so that there is no classical turning point.

For definiteness, suppose the ordering is: $-\infty\to$ peak $\to$
valley ...  valley $\to$ peak $\to +\infty$. Then
\begin{eqnarray}
\int_{-\infty}^{+\infty}  {|k'|\over k} dx 
&=&
\left|\ln\left[{k_{\mathrm peak}^1\over k_{-\infty}}\right]\right| +
\left|\ln\left[{k_{\mathrm valley}^1\over k_{\mathrm peak}^1}\right]\right| +
\cdots
\nonumber\\
&&\left|
\ln\left[{ k_{\mathrm peak}^n\over k_{\mathrm valley}^{n-1} }\right]
\right| +
\left|\ln\left[{k_{+\infty}\over k_{\mathrm peak}^n}\right]\right|.
\end{eqnarray}
Defining
\begin{eqnarray}
\Pi_p(k) &\equiv& \prod_{\mathrm peaks}k_{\mathrm peak}^i,
\\
\Pi_v(k)  &\equiv& \prod_{\mathrm valleys}k_{\mathrm valley}^i,
\\
\Pi_e(k)  &\equiv& \prod_{\mathrm extrema}k_{\mathrm extremum}^i,
\end{eqnarray}
we see
\begin{equation}
\int_{-\infty}^{+\infty}  {|k'|\over k} dx 
=
\left|\ln\left[
{\Pi_p^2(k) \over k_{-\infty} k_{+\infty} \Pi_v^2(k)}
\right]\right|.
\end{equation}
This bounds the Bogolubov coefficients as
\begin{equation}
|\alpha| \leq 
\cosh\left|\ln\left[
{\Pi_p(k)\over \sqrt{k_{-\infty} k_{+\infty}}\Pi_v(k)}
\right]\right|.
\end{equation}
That is
\begin{equation}
\label{B:alpha6}
|\alpha| \leq  
{
k_{-\infty}k_{+\infty} \Pi_{v}^2(k) + \Pi_{p}^2(k)
\over 
2\sqrt{k_{+\infty}k_{-\infty}} \Pi_{e}(k)
},
\end{equation}
and
\begin{equation}
\label{B:beta5}
|\beta| \leq  
{
|k_{-\infty}k_{+\infty} \Pi_{v}^2(k) - \Pi_{p}^2(k)|
\over 
2\sqrt{k_{+\infty}k_{-\infty}} \Pi_{e}(k)
}.
\end{equation}
Then the transmission and reflection probabilities satisfy
\begin{equation}
\label{B:T6}
T \geq 
{ 
4 k_{+\infty}k_{-\infty} \Pi_{e}^2(k)
\over 
\left\{  
\Pi_{p}^2(k) + k_{+\infty}k_{-\infty} \Pi_{v}^2(k)
\right\}^2},
\end{equation}
and
\begin{equation}
\label{B:R6}
R \leq 
{
\left\{  
\Pi_{p}^2(k) -k_{+\infty}k_{-\infty} \Pi_{v}^2(k)
\right\}^2 
\over 
\left\{  
\Pi_{p}^2(k) + k_{+\infty} k_{-\infty} \Pi_{v}^2(k)
\right\}^2}.
\end{equation}
In these formulae, peaks and valleys can be interchanged in the
obvious way, and by letting the initial or final peak sink down to
$V_{\pm\infty}$ as appropriate we obtain bounds for sequences such as:
$-\infty\to$ valley $\to$ peak ...  valley $\to$ peak $\to +\infty$,
or: $-\infty\to$ peak $\to$ valley ...  peak $\to$ valley $\to
+\infty$. In the case of one or zero extrema these formulae reduce to
the previously given results. [Equations
(\ref{B:T5-1})--(\ref{B:R5-1}).]  Further modifications of these
formulae are still possible, the cost is that more specific
assumptions are needed to derive more specific results.

%------------------------------------------------------------------------------
\section{Parametric Oscillations} 
%-----------------------------------------------------------------------------

Though the discussion so far has been presented in terms of the
spatial properties of the time-independent Schrodinger equation, the
mathematical structure of parametrically excited oscillations is
identical, needing only a few minor translations to be brought into
the current form. For a parametrically excited oscillator we have
\begin{equation}
{d^2 \phi\over dt^2} = \omega(t)^2 \phi.
\end{equation}
Just map $t\to x$, $\omega(t) \to k(x)$, and $\phi \to \psi$. In the
general analysis of equations (\ref{E:vartheta})--(\ref{B:R0}) the
quantity $\vartheta$ should be replaced by
\begin{equation}
\vartheta[\varphi(t),\omega(t)] \define
{\sqrt{(\varphi'')^2+ \left[\omega^2-(\varphi')^2\right]^2} 
\over 2 |\varphi'| }. 
\end{equation}
The analysis then parallels that of the Schrodinger equation.
Some key results are given below.

%---------------------------------------------------------------------
\subsection{\em Special Case 1}

If $\omega(-\infty) = \omega_0 = \omega(+\infty)\neq 0$, then by
choosing the auxiliary function to be $\varphi = \omega_0 t$ we can use
equations [(\ref{B:T1})--(\ref{B:R1})] to deduce
\begin{equation}
| \alpha | \leq 
\cosh\left( 
{1\over2\omega_0} \int_{-\infty}^{+\infty} |\omega^2(t)-\omega_0^2| dt  
\right),
\end{equation}
and
\begin{equation}
| \beta | \leq 
\sinh\left(
{1\over2\omega_0} \int_{-\infty}^{+\infty} |\omega^2(t)-\omega_0^2| dt 
\right).
\end{equation}

%---------------------------------------------------------------------
\subsection{\em Special Case 2}

If $\omega(-\infty)$ and $\omega(+\infty)\neq 0$ are both finite so
that suitable asymptotic states exist, and assuming $\omega^2(t)\geq
0$ so that the frequency is always positive, then applying equations
[(\ref{B:T2})--(\ref{B:R2})] to the case of parametric resonance
yields
\begin{equation}
| \alpha | \leq 
\cosh\left|
\int_{-\infty}^{+\infty} {|\omega'(t)|\over|\omega(t)|} dt  
\right|,
\end{equation}
and
\begin{equation}
| \beta | \leq 
\sinh\left|
\int_{-\infty}^{+\infty} {|\omega'(t)|\over|\omega(t)|} dt 
\right|.
\end{equation}

%---------------------------------------------------------------------
\subsection{\em Special case 2-a}

Suppose now that $\omega^2(t)$ is positive semidefinite, continuous,
and monotonic increasing or decreasing, varying from $\omega_{-\infty}
= \omega(-\infty)\neq 0$ to $\omega_{+\infty} = \omega(+\infty)\neq
0$.  The Bogolubov coefficients satisfy
\begin{equation}
|\alpha|  \leq 
{
\omega_{-\infty}  +\omega_{+\infty}
\over
2 \sqrt{\omega_{-\infty} \omega_{+\infty}}
},
\end{equation}
and
\begin{equation}
|\beta| \leq 
{
|\omega_{-\infty}  -\omega_{+\infty}|
\over
2 \sqrt{\omega_{-\infty} \omega_{+\infty}}
}.
\end{equation}

%---------------------------------------------------------------------
\subsection{\em Special Case 2-b}

Under the restriction $\omega(-\infty) = \omega_0 =
\omega(+\infty)\neq 0$, with the additional constraint that
$\omega(t)$ has a single unique extremum (either a maximum or a
minimum but not both), and provided that $\omega_{\mathrm
extremum}^2>0$ so that we do not encounter complex frequencies (no
classical turning point), the Bogolubov coefficients satisfy
\begin{equation}
|\alpha|  \leq 
{
\omega_0^2  +\omega_{\mathrm extremum}^2
\over
2 \omega_0 \omega_{\mathrm extremum}
},
\end{equation}
and
\begin{equation}
|\beta| \leq 
{
|\omega_0^2  -\omega_{\mathrm extremum}^2|
\over
2 \omega_0 \omega_{\mathrm extremum}
}.
\end{equation}

%---------------------------------------------------------------------

Suppose now that $\omega^2(t)$ has a single unique extremum (either
a peak or a valley), but that  $\omega(+\infty) \neq \omega(-\infty)$,
and further that $\omega^2(t) >0$ so that there is no classical
turning point. The Bogolubov coefficients satisfy
\begin{equation}
|\alpha|  \leq 
{
\omega_{-\infty}\omega_{+\infty}+\omega_{\mathrm extremum}^2
\over
2 \sqrt{\omega_{-\infty} \omega_{+\infty}} \omega_{\mathrm extremum}
},
\end{equation}
and
\begin{equation}
|\beta| \leq 
{
|\omega_{-\infty}\omega_{+\infty}-\omega_{\mathrm extremum}^2|
\over
2 \sqrt{\omega_{-\infty} \omega_{+\infty}} \omega_{\mathrm extremum}
}.
\end{equation}

%---------------------------------------------------------------------
\subsection{\em Special case 2-c}

Suppose now that $\omega(t)$ has a number of extrema (both peaks and
valleys). I allow $\omega(+\infty) \neq \omega(-\infty)$, but demand
that for all extrema $\omega_{\mathrm extremum}^i>0$ so that there is
no classical turning point.

For definiteness, suppose the ordering is: $-\infty\to$ peak $\to$
valley ...  valley $\to$ peak $\to +\infty$. Define
\begin{eqnarray}
\Pi_p(\omega)  &\equiv& \prod_{\mathrm peaks}\omega_{\mathrm peak}^i,
\\
\Pi_v(\omega)  &\equiv& \prod_{\mathrm valleys}\omega_{\mathrm valley}^i,
\\
\Pi_e(\omega)  &\equiv& \prod_{\mathrm extrema}\omega_{\mathrm extremum}^i.
\end{eqnarray}
The Bogolubov coefficients
satisfy
\begin{equation}
\label{B:alpha6b}
|\alpha| \leq  
{
\omega_{-\infty}\omega_{+\infty} \Pi_{v}^2(\omega) + \Pi_{p}^2(\omega)
\over 
\sqrt{\omega_{+\infty}\omega_{-\infty}} \Pi_{e}(\omega)
},
\end{equation}
and
\begin{equation}
\label{B:beta5b}
|\beta| \leq  
{
|\omega_{-\infty}\omega_{+\infty} \Pi_{v}^2(k) - \Pi_{p}^2(k)|
\over 
\sqrt{\omega_{+\infty}\omega_{-\infty}} \Pi_{e}(k)
}.
\end{equation}
In these formulae, peaks and valleys can be interchanged in the
obvious way, and by letting the initial or final peak sink down to
$\omega_{\pm\infty}$ as appropriate we obtain bounds for sequences such
as: $-\infty\to$ valley $\to$ peak ...  valley $\to$ peak $\to
+\infty$, or: $-\infty\to$ peak $\to$ valley ...  peak $\to$ valley
$\to +\infty$. In the case of one or zero extrema these formulae
reduce to the previously given results.

Again, further specializations of these formulae are still
possible. As always there is a trade-off between the strength of the
result and its generality.

%--------------------------------------------------------------------------
\section{Comparison with known analytic results} 
%--------------------------------------------------------------------------

For comparison purposes, in this section I collect several known
analytic results and show how they relate to the general results
presented in this article.

%----------------------------------------------------------------------------
\subsection{Delta-function potential}

For a delta function potential
\begin{equation}
V(x) = \alpha \; \delta(x),
\end{equation}
the transmission coefficient is known to be~\cite{Baym,Gasiorowicz}
\begin{equation}
T = {1\over 1 + {m \alpha^2\over2 E \hbar}}.
\end{equation}
This satisfies the bound (\ref{B:T1}), and also (\ref{B:T1-weak}), and
for $E\to\infty$ asymptotically approaches the bound, thus showing
that the bound cannot be improved in the high-energy regime {\em
unless additional hypotheses are made}.

Though these bounds were {\em derived} assuming well-behaved functions
the statements (\ref{B:T1}) and (\ref{B:T1-weak}) continue to make
good sense even for delta-function potentials. Thus any smooth set of
well-behaved functions tending to a delta-function limit may be used
to establish (\ref{B:T1}) and (\ref{B:T1-weak}) even for potentials
containing delta-function contributions.

%--------------------------------------------------------------------------
\subsection{Double-delta-function potential}

For the double delta function
\begin{equation}
V(x) = \alpha \left\{ \delta(x-L/2) + \delta(x+L/2) \right\},
\end{equation}
the transmission coefficient is~\cite{Galindo-Pascual}
\begin{equation}
T = { 1
\over 
1 + 
\left[ {2m\alpha\over\hbar^2 k} \cos(kL) + 
{1\over2}\left({2m\alpha\over\hbar^2 k}\right)^2 \sin(kL) \right]^2 
}.
\end{equation}
It is an easy exercise to check that this satisfies the bounds
(\ref{B:T1}) and (\ref{B:T1-weak}).

%----------------------------------------------------------------------------
\subsection{Square barrier} 

Tunneling {\em over} a square barrier is an elementary problem which
however is not always discussed in the textbooks.  (Tunneling {\em
under} a square barrier is much more popular.) The exact transmission
coefficient is
\begin{equation}
\label{E:T-square-well}
T = 
{ E(E-V_{e})
\over 
E(E-V_{e}) + {1\over4}V_{e}^2\sin^2(\sqrt{2m(E-V_{e})} L/\hbar)}.
\end{equation}
(See Landau-Lifshitz~\cite{Landau-Lifshitz} or Schiff~\cite{Schiff}.)
If we re-write this as
\begin{equation}
T = 
{ 
1
\over 
1 + {m V_{e}^2 L^2\over 2 E \hbar^2} 
{\sin^2(\sqrt{2m(E-V_{e})} L/\hbar)\over2 m (E-V) L^2/\hbar^2}
},
\end{equation}
then it is clear that the bound (\ref{B:T1}) is satisfied.  It is also
possible to verify that this satisfies the general lower bound
(\ref{B:T2}) that I have presented above, and in fact oscillates
between this lower bound and the upper $T\leq1$ unitarity limit.  For
certain values of the barrier width [$k_{\mathrm extremum} L = (2n+1)\pi/2$]
the square well saturates this bound thus showing that this bound
cannot be improved {\em unless additional hypotheses are made}.

%----------------------------------------------------------------------------
\subsection{Tanh potential}

For a smoothed step function of the form
\begin{equation}
V(x) = {V_{-\infty} + V_{+\infty} \over 2}
+ {V_{+\infty} - V_{-\infty} \over 2} \tanh\left({x\over L}\right),
\end{equation}
the reflection coefficient is known analytically to be~\cite{Landau-Lifshitz}
\begin{equation}
R = \left( {
\sinh[2\pi(k_{-\infty}-k_{+\infty})L]
\over
\sinh[2\pi(k_{-\infty}+k_{+\infty})L]
} \right)^2.
\end{equation}
This certainly satisfies the general bounds (\ref{B:T3})--(\ref{B:R3})
enunciated above, and as $L\to 0$ approaches and saturates the bound.

%----------------------------------------------------------------------------
\subsection{Sech potential}

For a $\sech^2$ potential of the form
\begin{equation}
V(x) = V_{e} \; \sech^2(x/L),
\end{equation}
the transmission coefficient is known analytically to
be~\cite{Landau-Lifshitz}
\begin{equation}
T = {
\sinh^2[\pi \sqrt{2mE} L/\hbar]
\over
\sinh^2 [\pi \sqrt{2mE} L/\hbar] + 
\cos^2[{1\over2}\pi\sqrt{1-8m V_{e} L^2/\hbar^2}]
},
\end{equation}
provided $8m V_{e} L^2 < \hbar^2$.  This satisfies the general
bounds, both (\ref{B:T0}) and (\ref{B:T3}), enunciated above.
(Though proving this is tedious.) Start by noting that for this
sech potential
\begin{equation}
T \geq \tanh^2[\pi \sqrt{2mE} L/\hbar],
\end{equation}
and use the inequality ($x>0$)
\begin{equation}
\tanh^2x > {x^2\over1+x^2} > \sech^2(1/x).
\end{equation}
Then
\begin{eqnarray}
T &\geq& \sech^2[\hbar/(\pi \sqrt{2mE} L)]
\\
&=& \sech^2\left[
{4\over\pi} \sqrt{m\over2E} 
{2L|V_{e}|\over\hbar} 
{\hbar^2\over8m|V_{e}|L^2}
\right].
\end{eqnarray}
Provided that the extremum is a peak, $V_{\mathrm peak}>0$ we can use
the bound $8m V_{\mathrm peak} L^2 < \hbar^2$ to deduce
\begin{equation}
T \geq  \sech^2\left[
\sqrt{m\over2E} 
{2L|V_{\mathrm peak}|\over\hbar} 
\right].
\end{equation}
This is the particularization of (\ref{B:T0}) to the present case.
If $V_{e} <0$ we need a different analysis.

%----------------------------------------------------------------------------
\subsection{Asymmetric square-well potential}

For the asymmetric square well
\begin{equation}
V(x) = \left\{ 
\matrix{
 V_1, \qquad x<a;\hfill\cr
 V_2, \qquad a<x<b;\hfill\cr
 V_3, \qquad b<x.\hfill}
\right.,
\end{equation}
we define $k_i\equiv\sqrt{2m(E-V_i)}/\hbar$. 
The transmission coefficient is~\cite{Messiah} 
\begin{equation}
T = { 4 k_1 k_2^2 k_3 
\over
(k_1+k_3)^2 k_2^2 + [k_1^2 k_3^2 + k_2^2(k_2^2-k_1^2-k_3^2)] \sin^2(k_2L)
}.
\end{equation}
Then
\begin{equation}
\label{B:asymetric-square-well}
T \geq { 4 k_1 k_2^2 k_3 
\over
(k_2^2 + k_1 k_3)^2
}.
\end{equation}
Similarly to the case for the symmetric square well, the transmission
probability for the asymmetric square well oscillates between the
bound (\ref{B:T5-1}) and the unitarity limit $T=1$. For certain values
of the width of the well [$k_2 L = (2n+1)\pi/2$] the transmission
coefficient saturates the bound thus showing that this bound cannot
be improved {\em unless additional hypotheses are made}. Because
$V_{-\infty}\neq V_{+\infty}$ the bound (\ref{B:T0}) is not
applicable, at least not without modification from its original
form.

%--------------------------------------------------------------------------
\subsection{Poschl-Teller potential}

For the Poschl-Teller potential
\begin{equation}
V(x) = V_0 \cosh^2\mu \left\{ \tanh([(x-\mu L)/L] + \tanh \mu \right\}^2,
\end{equation}
we have 
\begin{equation}
V_{-\infty} = V_0 e^{-2\mu}; \qquad 
V_{\mathrm extremum} = 0; \qquad
V_{-\infty} = V_0 e^{-2\mu}.
\end{equation}
The transmission coefficient is~\cite{Morse-Feshbach} 
\begin{eqnarray}
&&
T = 
\nonumber\\
&&
{
2 \sinh(\pi k_{-\infty} L) \sinh(\pi k_{+\infty} L)
\over
\cosh[\pi (k_{-\infty} + k_{+\infty}) L] + 
\cos\left[\pi\sqrt{1+{8mV_0 L^2\over\hbar^2}\cosh^2\mu}\right] 
}.
\nonumber\\
&&
\end{eqnarray}
It is now a straightforward if tedious exercise to check this analytic
result against all the bounds derived in this article.

%------------------------------------------------------------------------------
\section{Discussion} 
%-----------------------------------------------------------------------------

The various special cases discussed above are merely specific examples
of the general results (\ref{B:alpha0})--(\ref{B:beta0}) and
(\ref{B:T0})--(\ref{B:R0}) illustrating the power of the technique.
There are many other variations on the bounds presented above that can
be derived for specific choices of $\varphi(x)$ and specific
restrictions on the scattering potential $V(x)$.

The most general form of the bounds are given in equations
(\ref{B:alpha0})--(\ref{B:beta0}) and
(\ref{B:T0})--(\ref{B:R0}). Because of the large amount of freedom
in choosing the function $\varphi$ these bounds encode even more
specific cases beyond those discussed in this article, and have the
potential for leading to new interesting specific cases. The special
cases I discussed in this article were chosen for directness and
simplicity.

For instance, Special Case 1, as presented in equations
(\ref{B:T1})--(\ref{B:R1}) and (\ref{B:T1-weak})--(\ref{B:R1-weak}),
has the advantage that it applies to both scattering over the barrier
and under the barrier. On the other hand, Special Case 2, as presented
in equations (\ref{B:T2})--(\ref{B:R2}) and their specializations,
applies only to scattering over the barrier but has the advantage of
being much more selective in how much information is needed concerning
the scattering potential.

In summary, the bounds presented in this article are useful in
establishing {\em qualitative} analytic properties of one-dimensional
scattering, and as such are complementary to both explicit numerical
investigations and the guidance extracted from exact analytic
solutions.

%------------------------------------------------------------------------------
\section*{Acknowledgments} 
%-----------------------------------------------------------------------------

This research was supported by the US Department of Energy. I also
wish to thank LAEFF (Laboratorio de Astrof\'\i{}sica Espacial y
F\'\i{}sica Fundamental; Madrid, Spain) for hospitality during initial
phases of this research, and to acknowledge the kind hospitality of
Victoria University (Te Whare Wananga o te Upoko o te Ika a Maui;
Wellington, New Zealand) for hospitality during final stages of this
work.

%----------------------------------------------------------------------------

%----------------------------------------------------------------------------

\begin{references}
\bibitem{Transfer-matrix} 
Transfer matrix techniques are discussed, at
varying levels of detail, by Merzbacher~\cite{Merzbacher},
Singh~\cite{Singh}, and Mathews~\cite{Mathews}.
\end{references}
\end{document}